\documentclass[%
 reprint,
superscriptaddress,
 amsmath,amssymb,
 aps,
]{revtex4-2}

\usepackage{graphicx}
\usepackage{dcolumn}
\usepackage{bm}
\usepackage{hyperref}

\hypersetup{colorlinks=true, linkcolor=blue, citecolor=blue, filecolor=blue, urlcolor=blue}

\begin{document}


\title{Universal Scaling Behavior of Transport Properties in  Non-Magnetic RuO$_{2}$}
\author{Xin Peng}\thanks{These authors contribute equally.}
\affiliation{Department of Physics, China Jiliang University, Hangzhou 310018, China}
\author{Zhihao Liu}\thanks{These authors contribute equally.}
\affiliation{Beijing National Laboratory for Condensed Matter Physics and Institute of physics, Chinese academy of sciences, Beijing 100190, China}
\affiliation{University of Chinese academy of sciences, Beijing 100049, China}
\author{Shengnan Zhang}
\affiliation{Beijing Polytechnic College, Beijing 100042, China}
\author{Yi Zhou}
\affiliation{Department of Physics, China Jiliang University, Hangzhou 310018, China}
\author{Yuran Sun}
\affiliation{Department of Physics, China Jiliang University, Hangzhou 310018, China}
\author{Yahui Su}
\affiliation{Department of Physics, China Jiliang University, Hangzhou 310018, China}
\author{Chunxiang Wu}
\affiliation{School of Physics, Zhejiang University, Hangzhou 310027, China}
\author{Tingyu Zhou}
\affiliation{School of Physics, Zhejiang University, Hangzhou 310027, China}
\author{Le Liu}
\affiliation{School of Physics, Zhejiang University, Hangzhou 310027, China}
\author{Yazhou Li}
\affiliation{School of Physics, Hangzhou Normal University, Hangzhou 310036, China}
\author{Hangdong Wang}
\affiliation{School of Physics, Hangzhou Normal University, Hangzhou 310036, China}
\author{Jinhu Yang}
\affiliation{School of Physics, Hangzhou Normal University, Hangzhou 310036, China}
\author{Bin Chen}
\affiliation{School of Physics, Hangzhou Normal University, Hangzhou 310036, China}
\author{Yuke Li}
\affiliation{School of Physics, Hangzhou Normal University, Hangzhou 310036, China}
\author{Chuanying Xi}
\affiliation{Anhui Province Key Laboratory of Condensed Matter Physics at Extreme Conditions, High Magnetic Field Laboratory, Chinese Academy of Sciences, Hefei 230031, China}
\author{Jianhua Du}\email{Corresponding author:	jhdu@cjlu.edu.cn}
\affiliation{Department of Physics, China Jiliang University, Hangzhou 310018, China}
\author{Zhiwei Jiao}\email{jiaozw@cjlu.edu.cn}
\affiliation{Department of Physics, China Jiliang University, Hangzhou 310018, China}
\author{Quansheng Wu}\email{quansheng.wu@iphy.ac.cn}
\affiliation{Beijing National Laboratory for Condensed Matter Physics and Institute of physics, Chinese academy of sciences, Beijing 100190, China}
\affiliation{University of Chinese academy of sciences, Beijing 100049, China}
\author{Minghu Fang}\email{mhfang@zju.edu.cn}
\affiliation{School of Physics, Zhejiang University, Hangzhou 310027, China}
\affiliation{Collaborative Innovation Center of Advanced Microstructure, Nanjing 210093, China}


%


\date{\today}

\begin{abstract}
As a prototypical altermagnet, RuO$_{2}$ has been subject to many controversial reports regarding its magnetic ground state and the existence of crystal Hall effects. We obtained high-quality RuO$_{2}$ single crystal with a residual resistivity ratio (RRR = 152), and carefully measured its magnetization, longitudinal resistivity ($\rho_{xx}$) and Hall resistivity ($\rho_{yx}$) up to 35 T magnetic field. We also calculated its electronic band, Fermi surface, and conducted numerical simulations for its transport properties. It was found that no magnetic transition occurs below 400 K, and that all the transport properties are consistent with the numerical simulations results, indicating that the magnetotransport properties originate from the intrinsic electronic structures and are dominated by the Lorentz force. Particularly, no crystal Hall effects were observed in our RuO$_{2}$ samples and both magnetoresistance and Hall resistivity follow scaling behavior. These results demonstrate that RuO$_{2}$ is a typical semimetal, rather than an altermagnet.
\end{abstract}

\maketitle



Ruthenium dioxide (RuO$_{2}$) with rutile structure has long been considered as a Pauli paramagnetic metal \cite{ryden1970magnetic}, and has been used in  catalysis, microelectronics, and supercapacitors \cite{over2012surface,iles1967ruthenium,majumdar2019recent} due to its high catalytic activity and chemical stability. However, the neutron diffraction  and resonant x-ray scattering experiment seem to show that RuO$_{2}$ is an antiferromagnet with a high Néel temperature ($>$ 300 K) and a small magnetic moment of $\sim$0.05$\mu_{B}$/Ru \cite{PhysRevLett.118.077201,PhysRevLett.122.017202}. Subsequently, RuO$_{2}$ was considered as an altermagnet, a third fundamental magnetic phase beyond traditional ferromagnetism (FM) and antiferromagnetism (AFM) in crystals with collinear magnetic order \cite{feng2022anomalous,PhysRevLett.128.197202,bose2022tilted,PhysRevLett.129.137201,PhysRevX.12.040002,PhysRevX.12.031042,PhysRevX.12.040501,cuono2023orbital,jeong2024altermagnetic}, which has sparked much interest in its magnetism and transport properties. 

At first, it was predicted theoretically that the rutile RuO$_{2}$ would have the largest alermagnetic spin splitting, up to 1.4 eV \cite{PhysRevX.12.031042,PhysRevX.12.040501}, indicating its potential for various spintronics applications by proving a storage signal. However, very recently, the angle-resolved photoemission spectroscopy (ARPES) and spin-resolved ARPES (SARPES) performed on the thin film and single-crystal rutile RuO$_{2}$ did not detect the band splitting expected from altermagnetism \cite{PhysRevLett.133.176401}, indicating that RuO$_{2}$ is highly unlikely
to be an altermagnet. Second, as a model material of the collinear antiferromagnet, RuO$_{2}$ was expected to exhibit the spontaneous Hall effect due to its crystal symmetry breaking, so-called the crystal Hall effect (CHE), in which the crystal chirality can be used to control the sign of the Hall effect \cite{vsmejkal2020crystal}. However, both the muon spin rotation ($\mu$SR) \cite{PhysRevLett.132.166702} and the nuclear magnetic resonance (NMR) \cite{PhysRevB.60.12279} measurement demonstrate the absence of long-range magnetic order, supported by the density-functional theory (DFT) + $U$ calculation by A. Smolyanyuk  $et~ al.$ \cite{PhysRevB.109.134424}. In order to check whether the CHE emerges in RuO$_{2}$, a careful investigation of its transport properties is needed using high-quality single crystals. 
To confirm the presence of CHE in RuO$_{2}$ by the transport properties measurements becomes a direct evidence for its altermagnetism.

In this letter, we report on our work with high-quality RuO$_{2}$ single crystals, characterized by a residual resistivity ratio (RRR = 152). We meticulously measured their magnetization, longitudinal resistivity ($\rho_{xx}$), and Hall resistivity ($\rho_{yx}$), and calculated their electronic band and Fermi surface (FS) using  DFT. Additionally, we conducted numerical simulations for their transport properties. We found that no magnetic transition occurs below 400 K, indicating that our RuO$_{2}$ single crystals are paramagnetic, rather than collinear antiferromagnets. We also observed that the magnetoresistance (MR), $\rho_{yx}(B)$, and Hall conductivity $\sigma_{xy}(B)$, as well as the anisotropy of $\rho_{xx}$, are consistent with the results from numerical simulations, suggesting that the calculated bands can accurately describe all the transport properties of our RuO$_{2}$ crystals. Notably, no CHE was observed in our RuO$_{2}$ crystals. These findings further suggest that RuO$_{2}$ is unlikely to be an altermagnet.


RuO$_2$ single crystals were grown using a chemical vapor transport (CVT) method. Polycrystalline RuO$_2$ (99.9\%, Alfa Aesar) was sealed in an evacuated quartz tube with 10 mg/cm$^3$ TeCl$_4$ as a transport agent and then heated for two weeks at 1100 $^{\circ}$C in a tube furnace with a gradient 100 $^{\circ}$C. Polyhedral crystals with typical dimensions of $2\times0.5\times0.5$ mm$^{3}$ (see the inset of Fig. 1(b))were found in the cold end of the quartz tube. The composition of the crystals, confirmed as Ru : O = 1 : 2, was verified using the energy-dispersive x-ray (EDX) spectrometer. The crystal structure was determined using a powder x-ray diffractometer (XRD, PANalytical, Rigaku Gemini A Ultra) with the sample produced by grinding pieces of crystals. Electrical resistivity ($\rho_{xx}$), Hall resistivity ($\rho_{yx}$) were conducted using a Quantum Design physical property measurement system (PPMS-9 T) and a water-cooled magnet with the highest magnetic field up to 35 T, and magnetization measured on a Quantum Design magnetic property measurement system (MPMS-7 T). Meanwhile, the numerical simulations were carried out based on the Boltzmann transport theory and first-principles calculations, for comparing with the experimental results of $\rho_{xx}$ and $\rho_{yx}$. DFT calculations were performed using the Vienna ab initio simulation package (VASP) \cite{PhysRevB.54.11169,PhysRevB.59.1758} with the generalized gradient approximation (GGA) of Perdew, Burke, and Ernzerhof (PBE) for the exchange-correlation potential \cite{PhysRevLett.77.3865}. A cutoff energy of 520 eV and a $7\times7\times11$ k-point mesh were used to perform the bulk calculations. The Fermi surface (FS), magnetoresistance (MR) and Hall resistivity calculations were performed using the open-source software WannierTools \cite{WU2018405}, which is based on the Wannier tight-binding model (WTBM) \cite{PhysRevB.56.12847,PhysRevB.65.035109,RevModPhys.84.1419} constructed using Wannier90 \cite{MOSTOFI20142309}.


\begin{figure}[!htbpb]
	\includegraphics[width= 8.6cm]{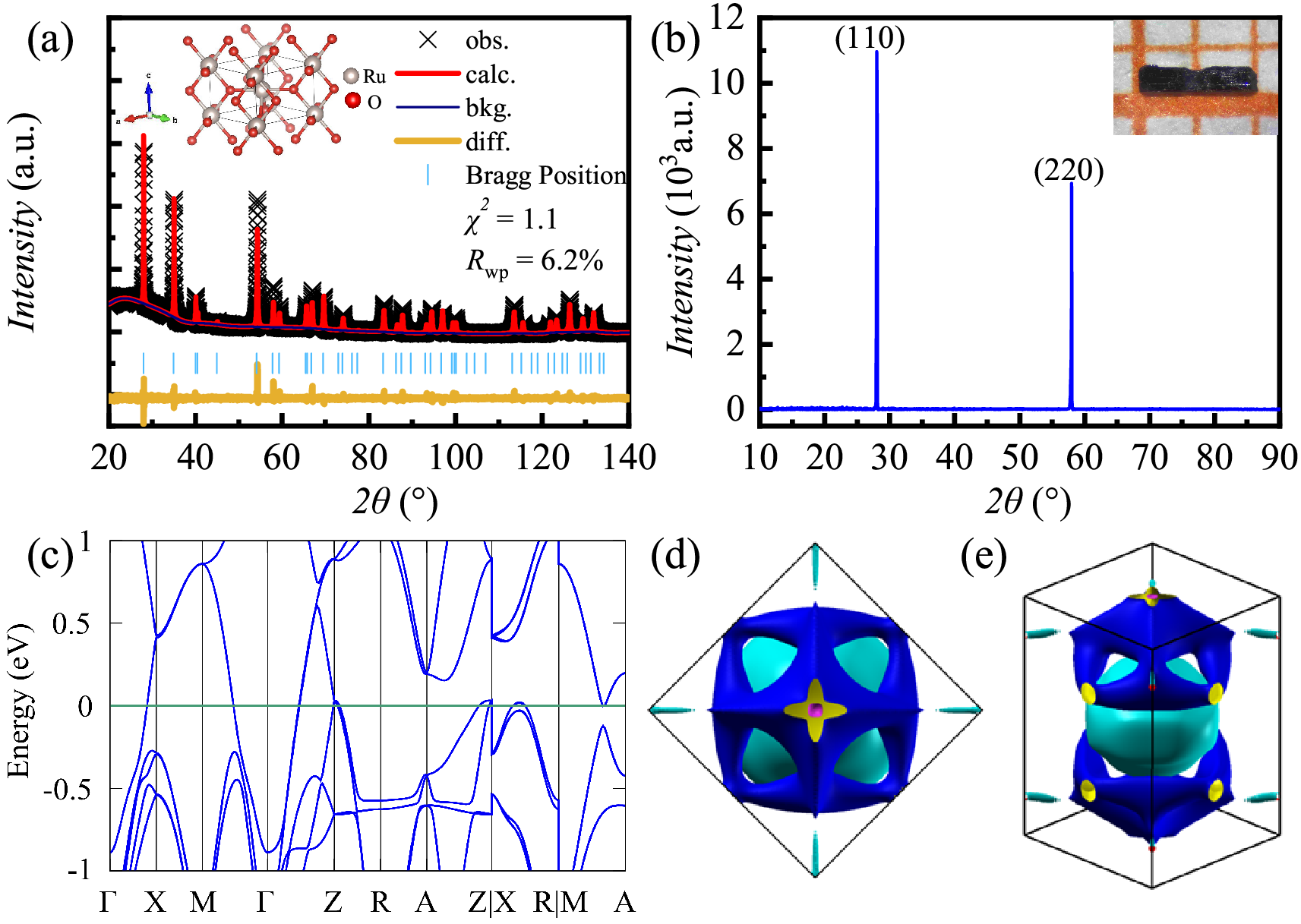}
      \caption{\label{FIG. 1}(Color online) (a) Polycrystalline XRD pattern with refinement profile at room temperature and crystal structure of RuO$_2$ (inset). (b) Single-crystal XRD pattern and photograph of RuO$_2$ crystal (inset). (c) Band structures with spin-orbit coupling (SOC). (d) and (e) Top and side views of FS calculated with SOC.}
\end{figure}

Figure 1(a) shows the powder XRD pattern, confirming that RuO$_{2}$ crystallizes in the rutile structure with nonsymmorphic space-group symmetry $P4_{2}/mnm$ (No.136). The lattice parameters $\textit{a} = \textit{b} = 4.501(2)$ \AA, and $\textit{c} = 3.114(3)$ \AA\ were determined using Rietveld refinement of the powder XRD data, yielding a weighted profile factor R$_{wp}$ of 6.2\% and a goodness of fit $\chi^{2}$ of 1.1. The primitive cell consists of two RuO$_{2}$ molecules, with two Ru atoms occupying the Wyckoff 2(\textit{a}) sites, and four O atoms occupying the 4(\textit{f}) sites, as illustrated in the inset of Fig. 1(a).  Each Ru atom is surrounded by a distorted octahedron of six O atoms. Proper rotation of the surrounding oxygen octahedra renders the two Ru positions—at the center and the corner—equivalent, indicating the coexistence of inversion symmetry ($\mathcal{P}$), time-reversal symmetry ($\mathcal{T}$), and nonsymmorphic symmetries in RuO$_{2}$. These symmetries impose strict constraints that hinder the stabilization of antiferromagnetic (AFM) ordering. $\mathcal{P}$ ensures the cancellation of magnetic moments, while $\mathcal{T}$ mandates invariance under time reversal, incompatible with a unique AFM state. Nonsymmorphic symmetries further complicate the formation of long-range AFM order. Based on the structure and lattice parameters detailed above, we perform DFT calculations of the band structure and FS. Figures 1(c), (d), and (e) show the band structure and FS with spin-orbit coupling (SOC) considered; for more details about the band and FS without SOC, see the Fig. S1 of Supplemental Material (SM). The band structure exhibits multiple bands crossing the Fermi level, indicating metallic behavior, and significant splitting due to SOC, especially near high-symmetry points like $\mathit{\Gamma}$ and \textit{X}, but this does not lead to the characteristic band splitting observed in altermagnets.  It has been suggested that potential Dirac points near high-symmetry points in the Brillouin zone, particularly around the \textit{X} and \textit{M} points, indicate the presence of topological nodal lines protected by crystal symmetries \cite{PhysRevB.95.235104,PhysRevB.98.241101}. The FS, shown from both top and side views, displays a complex, multi-sheeted structure with both electron and hole pockets; the electron pockets are prominently located near the $\mathit{\Gamma}$ point, while the hole pockets are distributed around the \textit{M} and \textit{A} points. The three-dimensional dispersion of these pockets shows significant $k_{z}$ dependence, typical in topological semimetals. The intricate FS topology supports the possibility of Dirac points, as evidenced by potential band intersections near the Fermi level, indicating regions of high Berry curvature. Notably, our electronic structure calculations, conducted within the framework of DFT, converge to a nonmagnetic solution without requiring additional adjustments.

\begin{figure}
	\includegraphics[width= 8.6cm]{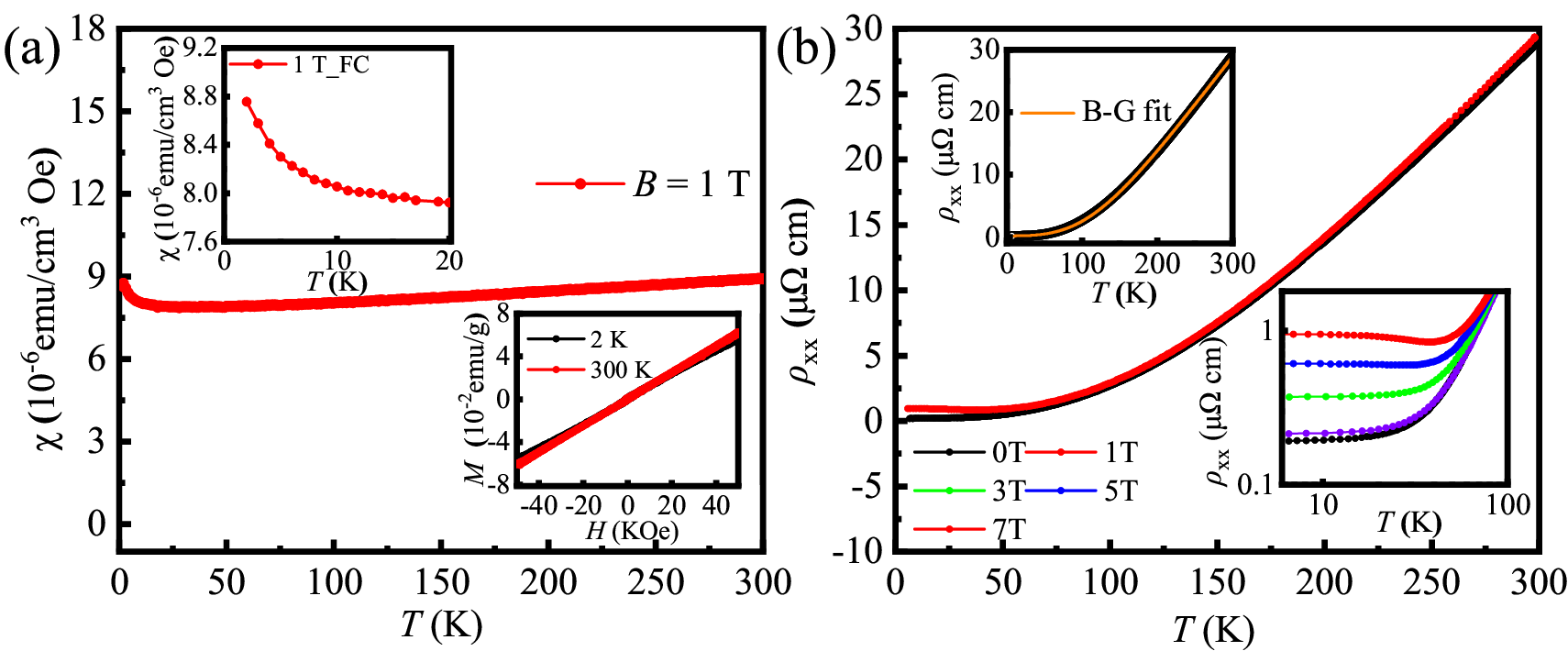}
	\caption{\label{FIG. 2}(Color online) (a) Temperature dependence of magnetization ($M$) measured at $\mu_0H = 1$ T perpendicular to the (110) plane with zero-field cooling (ZFC) processes. The left inset shows the temperature dependence of magnetic susceptibility below 20 K, and the right inset displays magnetization ($M$) as a function of magnetic field ($H$) measured at temperatures $T = 2$ K and $300$ K. (b) Temperature dependence of resistivity $\rho_{xx}$ at various magnetic fields. The left inset shows the fitting results, and the right inset shows an enlarged view at low temperatures.}  
\end{figure}

Figure 2(a) shows the temperature dependence of susceptibility measured at magnetic field $B$ = 1 T applied perpendicular to the (110) plane with a zero-field cooling (ZFC) process. The susceptibility decreases slightly with decreasing temperature, reaches a minimum around 25 K, and then increases. No magnetic transition was observed in entire temperature range (2 - 400 K), which is consistent with the recent observation by $\mu$SR and neutron diffraction studies \cite{PhysRevLett.132.166702,kessler2024absence}. The slight increase in susceptibility at low temperatures is attributed to the contribution of impurities containing Ru$^{3+}$ (\textit{s} = $\frac{5}{2}$) due to the presence of oxygen vacancies. As shown in the left inset of Fig. 2(a), the molar fraction of impurities was estimated to be 0.21($\pm$ 0.01) $\%$ by fitting the $\chi(T)$ data below 5 K using the modified Curie-Weiss law $\chi = \chi_{0} + \frac{C}{T-\theta}$. The linear dependence of magnetization ($M$) on magnetic field ($H$) [see the right inset of Fig. 2(a)] and the $\chi(T)$ behavior described above demonstrate that our RuO$_2$ crystal is paramagnetic with trace magnetic impurities.

Figure 2(b) presents the temperature dependence of the longitudinal resistivity $\rho_{xx}(T)$ for a RuO$_{2}$ single crystal (sample 1, S1). The current was applied in the (110) plane (see the inset of Fig. 1(b), the cleavage surface), with both zero field and a 7 T magnetic field applied perpendicular to the (110) plane.  The $\rho_{xx}$ measured at zero field decreases monotonically with decreasing temperature, with $\rho_{xx}$(300 K) = 28.87 $\mu\Omega$ cm and $\rho_{xx}$(6 K) = 0.19 $\mu\Omega$ cm. The residual resistivity ratio (RRR) = [$\rho_{xx}$(300 K)/$\rho_{xx}$(6 K)] = 152, indicating that our RuO$_{2}$ crystals are of high quality, surpassing those reported in previous studies (RRR = 80) \cite{PhysRevB.110.064432}. To elucidate the scattering mechanisms in RuO$_{2}$ crystals, we fitted the $\rho_{xx}(T)$ data using the Bloch-Grüneisen model \cite{ziman1960electrons}, as described by the equation: 
\begin{equation}
	\rho(T) = \rho_{0}+C(\frac{T}{\Theta_{D}})^{n}\int^{\Theta_{D}/T}_{0}\frac{x^{n}}{(e^{x}-1)(1-e^{-x})}dx
\end{equation}
where the residual resistivity $\rho_{0} = 0.19$ $\mu\Omega$ cm, the Debye temperature $\Theta_{D} = 900$ K, and the constant $C = 241.35$ $\mu\Omega$ cm were obtained from the fitting, as shown in the inset of Fig. 2(b). The power exponent $n = 2.73$ (close to 3) obtained from the best fit suggests that the scattering is influenced by phonon-assisted $s-d$ interband scattering \cite{ziman1960electrons, hu2020high, PhysRevB.92.180402}. Compared with $\rho_{xx}(T)$ measured at zero field, the $\rho_{xx}(T)$ at 7 T exhibits a similar behavior except for the enhancement of resistivity at low temperatures. As shown in the lower-right inset of Fig. 2(b), a field-induced upturn in resistivity, saturating to a field-independent constant, is observed, $i.e.$, extreme large magnetoresistance (XMR) emerging at low temperatures, similar to those observed in many topological trivial/non-trivial semimetals \cite{du2016large,PhysRevB.97.245101,PhysRevB.102.165133,PhysRevB.102.115145}.

\begin{figure*}
	\includegraphics[width= 17.2cm]{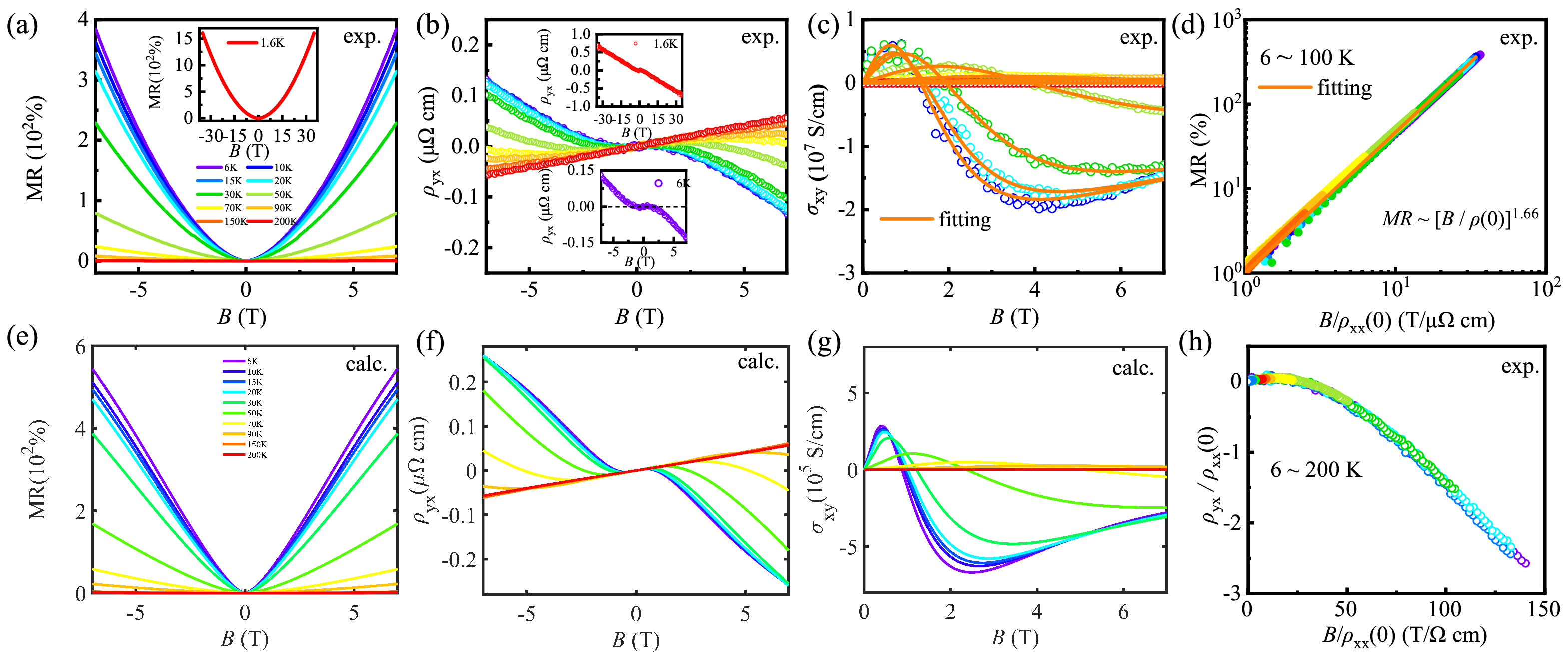}
	\caption{\label{FIG. 3} (Color online)(a) Measured magnetoresistance (MR) of RuO$_{2}$ at various temperatures, with an inset showing MR measured with a water-cooled magnet with fields up to 35 T at 1.6 K and 5 K. (b) Field dependence of Hall resistivity at various temperatures; the upper inset shows the Hall resistivity at 1.6 K and 5 K up to 35 T, while the lower inset shows the Hall resistivity at 6 K and 7 T. (c) Hall conductivity at various temperatures with a fitting line using a two-band model. (d) Kohler scaling analysis on MR with $m = 1.66$. (e), (f), (g) Calculated MR, Hall resistivity, and Hall conductivity from first-principles calculations at different temperatures. (h) Scaling analysis of Hall resistivity at various temperatures.}
   \end{figure*}

Next, we analyze the MR, Hall resistivity ($\rho_{yx}$), and Hall conductivity ($\sigma_{xy}$) experimental data using the scaling law, specifically the extended Kohler's law, and perform simulations based on Boltzmann transport theory and first-principles calculations. Considering the relaxation time approximation, the conductivity tensor $\sigma_{ij}$ can be calculated in the presence of a magnetic field by solving the Boltzmann equation, as described in \cite{Chambers1952, ashcroft1976solid, PhysRevB.99.035142,PhysRevResearch.6.043185,pi2024first,zhang2024in} and implemented in WannierTools~\cite{WU2018405}. 
 
\begin{equation}
	\sigma_{ij}^{(n)}(\bm{B}) = \frac{e^{2}}{\alpha\pi^{3}}\int d^{3}\bm{k}\tau_{n}v_{n}^{i}(\bm{k})\bar{v}_{n}^{j}(\bm{k})(-\frac{\partial f}{\partial \varepsilon})_{\varepsilon = \varepsilon _{n}(\bm{k})}
\end{equation}
where \textit{e} is the charge of electron, \textit{n} is the band index, $\alpha$ = 4 (or 8) when SOC is within (or without) in the Hamiltonian, $\tau_{n}$ is the relaxation time of $n$-th band, which is independent on the wave vector $\bm{k}$, \textit{f} is the Fermi-Dirac distribution,  $v_{n}^{i}(\bm{k})$ is the \textit{i}-th component of group velocities,
and $\bar{v}_{n}(\bm{k})$ is the weighted average of velocity over the past history of the charge carrier~\cite{ashcroft1976solid}. 
It's easy to derive that the conductivity tensor $\sigma$ divide  by $\tau$, $i.e.$, $\sigma$/$\tau$  is  a function of \textit{B}$\tau_{n}$
\begin{equation}
	\sigma_{ij}^{(n)}(\bm{B})/\tau_{n} = f(B\tau_{n})
\end{equation}
Assuming all bands have the same relaxation time $\tau_{n}$ = $\tau_{0}$,  the total conductivity $\sigma$($\bm{B}$) = $\Sigma$ $\sigma$$^{n}$($\bm{B}$) follows a scaling law: 
\begin{equation}
	\sigma(B)/\tau_{0} = g(B\tau_{0})
\end{equation}
According to the Drude model $\sigma_{0}$ = $\textit{ne}^{2}\tau_{0}/\textit{m}$, and $\rho$ = $\sigma^{-1}$, where \textit{n} is the carrier density and \textit{m} is the effective mass, we derive the Kohler's law \cite{Kohler1938,pippard1989magnetoresistance}:
\begin{equation}
	\Delta\rho/\rho_{0} = h(B/\rho_{0})
\end{equation}
 or the extended Kohler's law in its general form \cite{PhysRevX.11.041029}:
 \begin{equation}
 	\frac{\Delta\rho}{n\rho_{0}/m} = h(\frac{B}{n\rho_{0}/m})
 \end{equation}
The Hall resistivity could also exhibits scaling behavior \cite{zhang2024new}:
 \begin{equation}
	\frac{\rho_{yx}}{n\rho_{0}/m} = h(\frac{B}{n\rho_{0}/m})
\end{equation}

Figure 3(a) shows MR as a function of magnetic field $B$ at various temperatures, defined by the standard equation: MR = $\frac{\Delta\rho}{\rho(0)} = \left[\frac{\rho(B,T) - \rho_0(T)}{\rho_0(T)}\right] \times 100\%$, where $\rho(B,T)$ and $\rho_0(T)$ are the resistivities measured at field $B$ and zero field, respectively. The measured MR is extremely large at low temperatures, reaching $3.8 \times 10^{2}\%$ at 6 K and 7 T, which is nearly twice as large as that reported recently for RuO$_2$ single crystals at 2 K in 9 T \cite{PhysRevB.110.064432}, and it does not show any signs of saturation at the highest field of our measurement. To examine the MR behavior under high magnetic fields, measurements were performed with a water-cooled magnet up to 35 T. The inset of Fig. 3(a) shows the MR as a function of field up to 35 T at 1.5 and 5 K, where the MR reaches $1.5 \times 10^3\%$ at 1.5 K and 35 T and also shows no saturation. Figure 3(e) presents the numerical simulation results of MR as a function of $B$, based on Fermi surface calculations using first-principles, and experimental $\rho_0(T)$, which capture most of the essential features of the experimental MR curves, exhibiting a similar increase with field and a decrease with increasing temperature. The consistency between experimental and theoretical results demonstrates the validity of our first-principles approach in capturing the essential physics of MR in RuO$_2$. As discussed above, MR can be described by Kohler's scaling law \cite{pippard1989magnetoresistance}:
\begin{equation}
	\text{MR} = \frac{\Delta\rho_{xx}(T,H)}{\rho_{0}(T)} = \alpha(B/\rho_{0}(T))^{m}
\end{equation}
As show in the inset in Fig. 3(d), all MR data from \textit{T} = 6 K to 100 K collapse onto a single line when plotted as MR $\sim$ B/$\rho_{0}$ curve, with $\alpha$ = 0.01 ($\mu$$\Omega$ cm/T)$^{1.66}$ and \textit{m} = 1.66 obtained by fitting. The power field dependence of MR observed in RuO$_{2}$ is attributed to the Fermi surface structure discussed above. The MR data obtained from other sample (S2) measured up to 14 T exhibit a similar behavior (see Fig. S2 in the SM)

As one of altermagnetic candidates, RuO$_{2}$ would exhibit CHE due to its conventional time-reversal symmetry breaking in the electronic structure. We rechecked the possibility of CHE emerging in RuO$_{2}$ crystals through both experiments and theoretical analysis. Figure 3(b) shows the Hall resistivity ($\rho_{yx}$) as a function of magnetic field (\textit{B}) measured at various temperatures. As shown in the lower inset of Fig. 3(b), at 6 K,  $\rho_{yx}$ starts as a positive value, increases to a maximum, and then decreases to negative with increasing magnetic field. We also measured $\rho_{yx}$ at lower temperatures down to 1.6 K and higher fields up to 35 T, as shown in the upper inset of Fig. 3(b), exhibiting a similar behavior in the lower field region ($B < 7$ T). Below 70 K, all the $\rho_{yx}$ exhibit the same behavior with that at 6 K. At higher temperatures (90, 150, 200 K), $\rho_{yx}$ exhibits a linear behavior with a positive slop, indicating that hole carriers dominate. Then, we performed numerical simulations for the $\rho_{yx}(B)$. Similar to the MR data discussed above, based on the band structure and Boltzmann transport equation, as shown in Fig. 3(f), the simulations capture the essential behaviors of the measured $\rho_{yx}(B)$, especially the $\rho_{yx}(B)$ sign reversal behavior at lower temperatures. As shown in Fig. 3(h), it is found  that all $\rho_{yx}(B)$ data measured at different temperatures (6 - 100 K), normalized by the $\rho_{0}$ at each temperature, collapse onto a single curve, indicating that a scaling law can describe the $\rho_{yx}(B)$ behavior. Meanwhile, we calculated the Hall conductivity based on a two-band model using the $\rho_{xx}(B)$ and $\rho_{yx}(B)$ data measured at various temperatures by the equation \cite{PhysRevB.94.121101}:
\begin{equation}
	\sigma_{xy} = \frac{\rho_{yx}}{\rho_{xx}^{2}+\rho_{yx}^{2}} = eB[\frac{n_{h}\mu_{h}^{2}}{1+\mu_{h}^{2}B^{2}}-\frac{n_{e}\mu_{e}^{2}}{1+\mu_{e}^{2}B^{2}}]
\end{equation}
where $\textit{n}_{e}$($\textit{n}_{h}$) is the charge density of electrons (holes), $\mu_{e}$($\mu_{h}$) is the mobility of electrons (holes), and \textit{e} is the charge of electron. The calculated $\sigma_{xy}$ [see Fig. 3(g)] from the simulations can also capture the main behaviors of $\sigma_{xy}(B)$ measured at various temperatures. The $\textit{n}_{e}$, $\textit{n}_{h}$ and $\mu_{e}$, $\mu_{h}$ values at different temperatures were obtained by the simultaneously fitting to the experimental $\sigma_{xy}(B)$ data [see Fig. 3(c)] $\sigma_{xx}(B)$ data (see Fig. S4(a) in SM), such as at 6 K $\textit{n}_{e}$ = 9.27 $\times$ 10$^{21}$cm$^{-3}$, $\textit{n}_{h}$ = 8.47 $\times$ 10$^{21}$cm$^{-3}$, which are close but not completely compensated, similar to the case discussed in Ref \cite{zhang2024new}. 

\begin{figure}
	\includegraphics[width= 8.6cm]{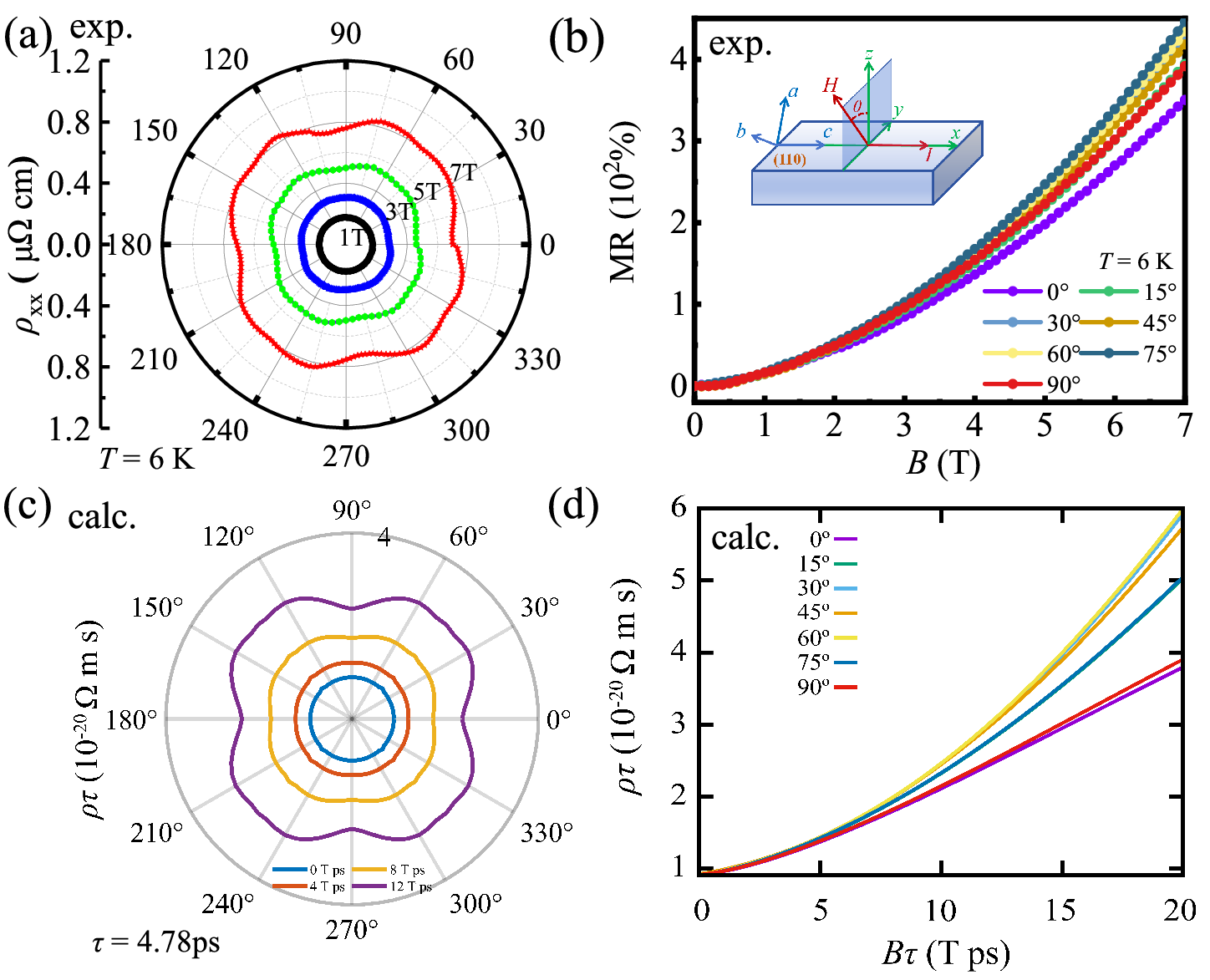}
	\caption{\label{FIG. 4}(Color online) (a), (c) Measured and calculated anisotropy of resistivity $\rho_{xx}$ for field rotated in \textit{a}-\textit{b} plane. (b), (d) Measured and calculated MR as a function of field for the various directions, inset of (b) Schematic diagram of resistivity measurements: current align \textit{c} axis, $\theta$ is the angle between field and \textit{z} axis.}
\end{figure}

At the same time, we measured the angular dependence of longitudinal resistivity $\rho_{xx}$ [see Fig. 4(a)], and MR [see Fig. 4(b)] in the \textit{ab} plane (with the applied current along the \textit{c}-axis), and performed numerical calculations for them, as shown in Figs. 4(c) and 4(d), respectively. Notably, $\rho_{xx}(B)$ exhibits inversion symmetry, $i.e.$, $\rho_{xx}(\theta) = \rho_{xx}(\theta + \pi)$, although its anisotropy is not significant, with a minimum at $\theta = 0^\circ$ when \textit{B} is applied in the [110] direction and a maximum at $\theta = 45^\circ$, corresponding to when $\theta$ aligns with the \textit{a/b}-axis. This behavior is consistent with the crystal structure of RuO$_{2}$ and the symmetry of the FS projected onto the \textit{ab} plane, as illustrated in Figs. 1(a), 1(d), and 1(e). A similar power-law field dependence of MR at 6 K, characterized as $\propto B^m$ (with \textit{m} ranging from 1.66 to 1.72 for different $\theta$ values), was confirmed by both measurements and numerical simulations, as shown in Figs. 4(b) and 4(d), respectively. Meanwhile, We also measured the resistivity $\rho_{xx}$ at 6 K with various field orientations relative to the applied current $I$, (see the inset of Fig. S3 in the SM), further confirming that the MR is determined by the Lorentz force on the carriers.

In summary, we successfully grew RuO$_{2}$ single crystals with rutile structure and performed the measurements of magnetization, longitudinal resistivity, and Hall resistivity. The numerical simulations for its transport properties were conducted based on Boltzmann transport theory and first-principles calculations. It was found that no magnetic transition occurs below 400 K, and all the transport properties can be described by the intrinsic electric structure and dominated by the Lorentz force, consistent with  numerical simulations results. Notably, no CHE was observed in our crystals. These results further demonstrate that RuO$_{2}$ is a typical semimetal, rather than an altermagnet.

\begin{acknowledgments}
 This research is supported by the National Key R$\&$D program of China under Grant No. 2022YFA1403202, 2023YFA1607400, 2024YFA1408400; The National Natural Science Foundation of China (Grant No. 12074335, 52471020, 12274436, 12188101); The Hangzhou Joint Fund of the Zhejiang Provincial Natural Science Foundation of China (under Grants No. LHZSZ24A040001). A portion of this work was performed on the Steady High Magnetic Field Facilities, High magnetic Field Laboratory, CAS.
\end{acknowledgments}

\nocite{*}

%

\end{document}